\documentclass[a4paper]{spie}  
\usepackage[]{graphicx}
\usepackage{wrapfig}

\title{Development and Performance of Kyoto's X-ray Astronomical SOI pixel (SOIPIX) sensor}

\author{
	Takeshi G. Tsuru\supit{a},	
	Hideaki Matsumura\supit{a},	
	Ayaki Takeda\supit{a},		
	Takaaki Tanaka\supit{a},	
	Shinya Nakashima\supit{b},	
	Yasuo Arai\supit{c},		
	Koji Mori\supit{d},			
	Ryota Takenaka\supit{d},	
	Yusuke Nishioka\supit{d},	
	Takayoshi Kohmura\supit{e},	
	Takaki Hatsui\supit{f,g}, 	
	Takashi Kameshima\supit{g},	
	Kyosuke Ozaki\supit{f},		
	Yoshiki Kohmura\supit{f},	
	Tatsuya Wagai\supit{f},		
	Dai Takei\supit{f},			
	Shoji Kawahito\supit{h},	
	Keiichiro Kagawa\supit{h},	
	Keita Yasutomi\supit{h},	
	Hiroki Kamehama\supit{h},	
	and
	Sumeet Shrestha\supit{h}	
\skiplinehalf
\supit{a}Department of Physics, Graduate School of Science, Kyoto University, \\
	Kitashirakawa, Sakyo-ku, Kyoto, 606-8502, Japan; \\
\supit{b}Institute of Space and Astronautical Science (ISAS)/JAXA, \\
	3-1-1 Yoshinodai, Chuo-ku, Sagamihara, Kanagawa 252-5210, JAPAN\\
\supit{c}Institute of Particle and Nuclear Studies, High Energy Accelerator Research Org., KEK, \\
	1-1 Oho, Tsukuba 305-0801, Japan\\
\supit{d}Department of Applied Physics, Faculty of Engineering, University of Miyazaki, \\
	1-1 Gakuen Kibana-dai Nishi, Miyazaki 889-2192, Japan\\
\supit{e}Department of Physics, Faculty of Science and Technology,
	Tokyo University of Science, \\
	2641 Yamazaki, Noda, Chiba 278-8510, Japan\\
\supit{f} RIKEN SPring-8 Center, 1-1-1 Kouto, Sayo-cho, Sayo-gun, Hyogo 679-5148, Japan\\
\supit{g} JASRI, 1-1-1 Kouto, Sayo-cho, Sayo-gun, Hyogo 679-5198, Japan\\
\supit{h} Research Institute  of Electronics, Shizuoka University, \\
	Johoku 3-5-1, Naka-ku, Hamamatsu 432-8011, Japan
}

\authorinfo{Send correspondence to T.G.T.: E-mail: tsuru@cr.scphys.kyoto-u.ac.jp}

\begin{document}
  \maketitle

\begin{abstract}
We have been developing monolithic active pixel sensors, known as Kyoto's X-ray SOIPIXs,
based on the CMOS SOI (silicon-on-insulator) technology for next-generation X-ray astronomy satellites.
The event trigger output function implemented in each pixel offers microsecond time resolution
and enables reduction of the non-X-ray background that dominates the high X-ray energy band above 5--10 keV.
A fully depleted SOI with a thick depletion layer and back illumination offers wide band coverage of
0.3--40 keV.  Here, we report recent progress in the X-ray SOIPIX development.
In this study, we achieved an energy resolution of 300~eV (FWHM) at 6~keV and a read-out
noise of 33~e- (rms) in the frame readout mode, which allows us to clearly resolve Mn-K$\alpha$ and
K$\beta$. Moreover, we produced a fully depleted layer with a thickness of $500~{\rm \mu m}$.
The event-driven readout mode has already been successfully demonstrated.
\end{abstract}



\section{INTRODUCTION}
\label{sec:intro}  
\begin{figure}
 \begin{center}
   \begin{tabular}{c}
   \includegraphics[width=14cm]{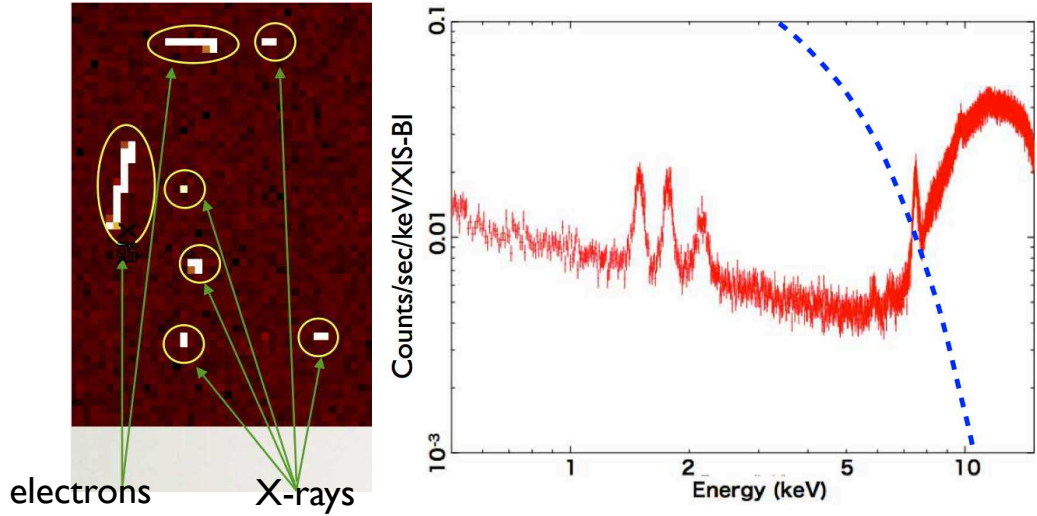}
   \end{tabular}
 \end{center}
   \caption[Figure_1]
   {\label{Figure_1}
	(Left) Raw frame data obtained with XIS of the Suzaku satellite on orbit.
	(Right) The NXB spectrum of Suzaku XIS-BI remaining
	after the NXB rejection\cite{2007KoyamaPASJ_Suzaku_XIS,2008TawaPASJ_XIS_NXB}.
	The dashed blue line shows an example of a diffuse source with the surface brightness of
	$1\times 10^{-14} {\rm ergs\ s^{-1} cm^{-2} arcmin^{-2}}$ at 2--10~keV and a photon index of $\Gamma=2.0$.
	}
\end{figure}

The X-ray charge-coupled device (CCD) is the current standard imaging spectrometer
for X-ray astronomy satellites at 0.3--10~keV.
They offer Fano limited spectroscopy with a readout noise better than 6~e- (rms),
with wide, fine imaging using a sensor size of $\sim$20--30~${\rm mm}$ and a pixel size of $\sim 24~{\rm \mu m}$.
However, X-ray CCDs have some weaknesses.
In particular, one of its serious weaknesses is high non-X-ray background (NXB),
which consists of particle events that cannot be distinguished from X-ray events.
The left panel of Figure~\ref{Figure_1} shows a raw image obtained with the X-ray imaging spectrometer (XIS)
on board the Suzaku satellite \cite{2007KoyamaPASJ_Suzaku_XIS}.
While a particle event usually produces a charge track,
an X-ray event has a small charge spread with one pixel or a couple of adjacent pixels.
Then, we usually reject NXB events with the charge spread.
The right panel of Figure~\ref{Figure_1} shows the remaining NXB spectrum after the rejection,
which is collected with entire field of view of the XIS back-illuminated (BI) CCD on board
Suzaku\cite{2007KoyamaPASJ_Suzaku_XIS,2008TawaPASJ_XIS_NXB}.
In the high-energy band, NXB is too high to observe astrophysically important diffuse sources such as
hard X-ray emission from supernova remnants and clusters of galaxies.
Thus, we started developing an SOI pixel sensor for future X-ray astronomy
to realize low NXB and extend the frontiers of wide-band imaging spectroscopy.
In this study, we briefly describe the status and achievements of this development.

\section{XRPIX: SOIPIX for X-ray Astronomy}
\begin{figure}
   \begin{center}
   \begin{tabular}{c}
   \includegraphics[width=14cm]{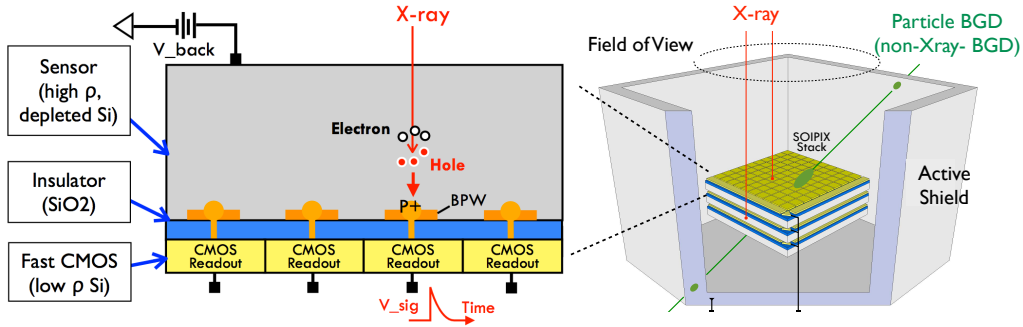} 
   \end{tabular}
   \end{center}
   \caption[Figure_2]
   {\label{Figure_2}
	(Left) The schematic view of the XRPIX cross-section.
	The X-ray is illuminated from the back.
	(Right) Conceptual drawing of the low non-X-ray background (BGD) detector with XRPIX and active shield.
	}
   \end{figure}

\begin{table}[b]
\caption{\label{tab:fonts}Target specification}
\vspace{-0.25cm}
\begin{center}
\begin{tabular}{ll}
\hline
\rule[-1ex]{0pt}{3.5ex}  Item & Specification  \\
\hline
\rule[-1ex]{0pt}{3.5ex}	Imaging & Chip size: $25\times 25~{\rm mm^2}$, Pixel Size: 30--60${\rm \mu m}$ \\
\rule[-1ex]{0pt}{3.5ex}	Bandpass & 0.3--40~keV \\
\rule[-1ex]{0pt}{3.5ex}	Energy resolution & $140~{\rm eV}$  (FWHM) at 6~keV \\
\rule[-1ex]{0pt}{3.5ex}	Readout noise & $10~{\rm e}$ (Requirement), $3~{\rm e}$ (Goal) (rms) \\
\rule[-1ex]{0pt}{3.5ex}	Time & Resolution: $10~{\rm \mu sec}$, Throughput: $2~{\rm kHz}$ \\
\rule[-1ex]{0pt}{3.5ex} Non-X-ray background & $1/100$ of X-ray CCD at 20~keV
						(${\rm 5\times 10^{-5}~cps/keV/10\times 10mm^{2}}$) \\
\hline
\end{tabular}
\end{center}
\end{table}
The SOI pixel sensor is monolithic using a low-resistivity Si bonded wafer for high-speed CMOS circuits,
an SiO$_2$ insulator, and a high-resistivity depleted Si layer for X-ray detection.
Our SOI pixel sensor for X-ray astronomy is referred to ``XRPIX'' (meaning, X-Ray PIXel),
whose target specification is shown in Table 1.
The goal of this development is the realization of a new X-ray detector system
having a much faster readout and lower NXB than X-ray CCDs
simultaneously with CCD equivalent-imaging spectroscopic performance.
To achieve low NXB, we apply the anti-coincidence technique
with scintillators surrounding XRPIX, as shown in Figure~\ref{Figure_2}.
We equip XRPIX with functions of the output trigger signal and the corresponding hit pixel address.
If the surrounding scintillators output a signal coincidentally with a trigger signal from XRPIX,
the event is rejected as a particle background event (anti-coincidence).
Only events that are not rejected by the anti-coincidence are read out from XRPIX.
Because of the kilohertz event rate from the surrounding scintillators,
XRPIX need to have time resolution much faster than kilohertz.
The anti-coincidence is estimated to reduce NXB of XRPIX
to the same level as that of the hard X-ray imager (HXI) of ASTRO-H\cite{2012KokubunSPIE_ASTRO-H_HXI},
which is two orders of magnitude lower than that of the X-ray CCD at 20~keV.

\begin{figure}
  \begin{center}
   \begin{tabular}{c}
   \includegraphics[width=10cm]{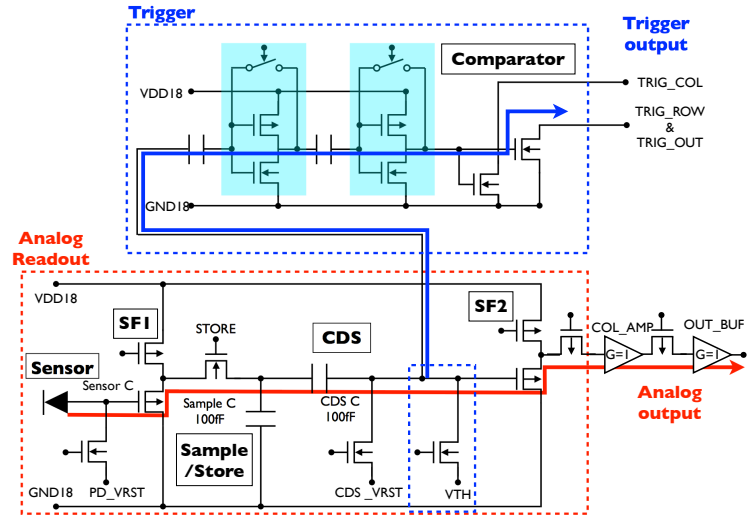}
   \end{tabular}
 \end{center}
   \caption[Figure_3]
   {\label{Figure_3}
	The CMOS circuit implemented in each pixel of XRPIX1.
	}
\end{figure}

Figure~\ref{Figure_3} shows the CMOS circuit implemented in each pixel of XRPIX1,
which is the first test device\cite{2011RyuIEEE_SOIPIX_XRPIX1-CZ-FI}.
The same essential pixel circuit has been used throughout the development in spite of minor revisions.
The red line shows the path of the analog signal.
The signal charge collected from the sensor region through the sense node is amplified by the source follower
labeled ``SF1.''
The reset noise generated at the sense node is canceled
by the in-pixel correlated double sampling circuitry labeled with ``CDS.''
The analog signal is read out through amplifiers with the second source follower labeled ``SF2,''
followed by the column amplifier (COL\_AMP) and output buffer (OUT\_BUF).
The blue line in the figure shows the path of the trigger signal,
which is generated by the circuit consisting of a two-stage chopper-type comparator
with a threshold set using the circuitry labeled with ``VTH.''

Because the low-noise pixel circuit development is the most important and difficult,
we have been focusing mainly on the pixel circuit, layout, and structure.
We have done frequent processes of small test devices with
imaging areas and chip sizes of 1--4.5~mm and 2.4--6.0~mm, respectively.
We have manufactured six text element groups (TEGs) since 2010 when we started the development of XRPIX.
The first model with a trigger-output function is XRPIX1\cite{2010RyuIEEE-NSS10_XRPIX1-CZ-FI,
2011RyuIEEE_SOIPIX_XRPIX1-CZ-FI, 2011RyuIEEE-NSS11_XRPIX1_EventDriven, 2012ShinyaPhysProc_XRPIX1-FZ-FI_TIPP2011}.
In XRPIX1b, the gain is increased by reducing the size of the buried p-well (BPW)
(see the next section)\cite{2013RyuIEEE_XRPIX1b-CZ-FI_SoftXray_CrossTalk, 2013TakedaIEEE_XRPIX1b-CZ-FI_EventDriven_SORMA_WEST, 2014MatsumuraNIMA_CCE_XRPIX1_1b_PIXEL2013}.
In XRPIX2, we increased the pixel size to $60~{\rm \mu m}$
and introduced multiple nodes\cite{2013ShinyaNIMA_XRPIX2-CZ-FI_PIXEL2012}, while XRPIX2b is the first buttable chip.
Finally, XRPIX3 and XRPIX3b are devices equipped with in-pixel charge sensitive amplifiers\cite{2013TakedaIEEE_NSS13_XRPIX3_33e_300eV}.
We also fabricated a TEG for in-chip analog-to-digital converter (ADC)\cite{2011ShinyaIEEE-NSS11_XRPIX-ADC1}.

\section{Spectral Performance in the Frame Readout Mode}
\label{sec:spectrum_frame_readout}
\begin{wrapfigure}{r}{7cm}
\vspace{-0.5cm}
 \begin{center}
   \begin{tabular}{c}
   \includegraphics[width=7cm]{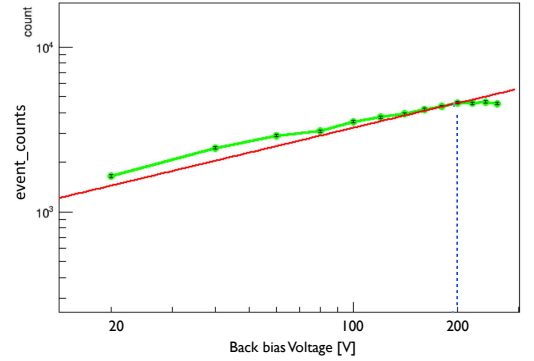}
   \end{tabular}
   \caption[Figure_4]
   {\label{Figure_4}
	The number of 22-keV X-ray events detected with the XRPIX1b
	high-resistivity floating zone wafer as a function of back-bias voltage.
	The data is adapted from\cite{2014MatsumuraNIMA_CCE_XRPIX1_1b_PIXEL2013}.
	The red line shows the counts proportional to the square of the back-bias voltage.
	}
 \end{center}
\vspace{-0.2cm}
\end{wrapfigure}
We show some frame mode performance results in which we did not use the trigger function. 
The entire sensor region was serially read out like a CCD.
The thick depletion layer is necessary to have high quantum efficiency for high-energy X-rays
(e.g., $>10~{\rm keV}$).
We are testing a floating zone (FZ) wafer with the high resistivity of $\sim$4--7~${\rm k\Omega}\cdot{\rm cm}$
(specification  $> 1~{\rm k\Omega}\cdot{\rm cm}$ )
as well as a standard Czochralski (CZ) wafer with $\sim 0.7~{\rm k\Omega}\cdot{\rm cm}$.
Figure~\ref{Figure_4} shows the count rate of 22-keV X-rays as a function of back-bias voltage
for an XRPIX1b equipped with the FZ-wafer\cite{2014MatsumuraNIMA_CCE_XRPIX1_1b_PIXEL2013}.
Because the attenuation length of $1200~{\rm \mu m}$ at 22 keV is longer than the physical wafer
thickness of $500~{\rm \mu m}$, the chip is optically thin for the X-rays.
The count rate increases as the back-bias voltage increases,
following the expected slope of the depletion layer thickness as a function of
the back-bias voltage before 200~V, and becomes constant above 200~V.
This result suggests that the full depletion of $500~{\rm \mu m}$ is reached at 200~V.
In addition to the thick depletion layer,
we also need back-illumination to have high quantum efficiency for low energy X-rays (e.g., $<1{\rm keV}$).
We boost the development of thin dead-layer on the back,
utilizing our experience from X-ray CCD developments\cite{2013TsunemiSPIE_ASTRO-H_SXI}.

\begin{figure}[b]
\begin{center}
 \begin{minipage}{0.45\hsize}
   \begin{center}
   \begin{tabular}{c}
   \includegraphics[height=6cm]{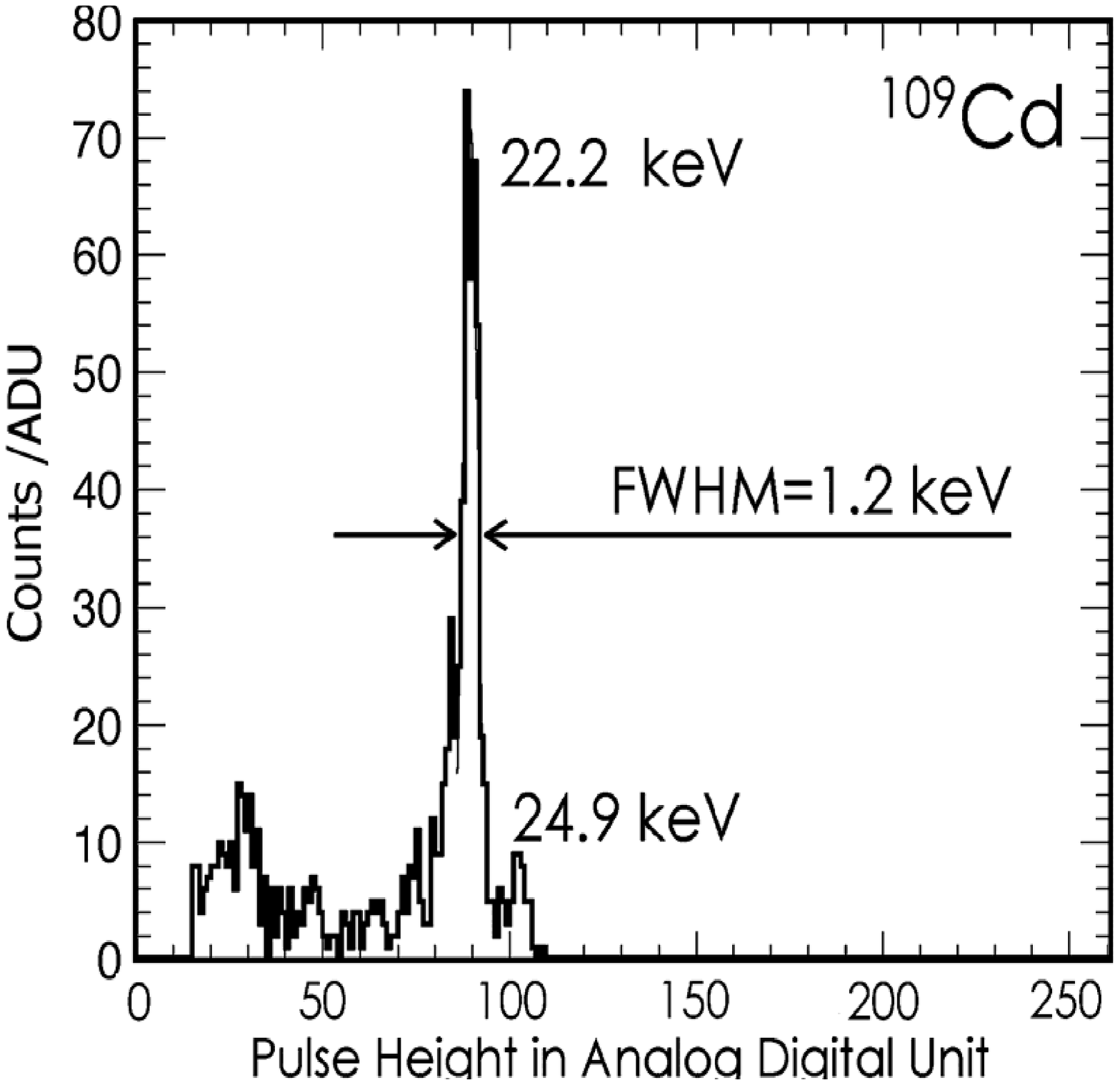}
   \end{tabular}
   \end{center}
   \caption[Figure_5]
   {\label{Figure_5}
	The Cd-109 spectrum obtained with XRPIX1 \cite{2011RyuIEEE_SOIPIX_XRPIX1-CZ-FI}, where
	one analog digital unit (ADU) corresponds to $224~\mu{\rm V}$.
	}
 \end{minipage}
 \hspace{0.05\hsize}
 \begin{minipage}{0.45\hsize}
   \begin{center}
   \begin{tabular}{c}
   \includegraphics[height=5cm]{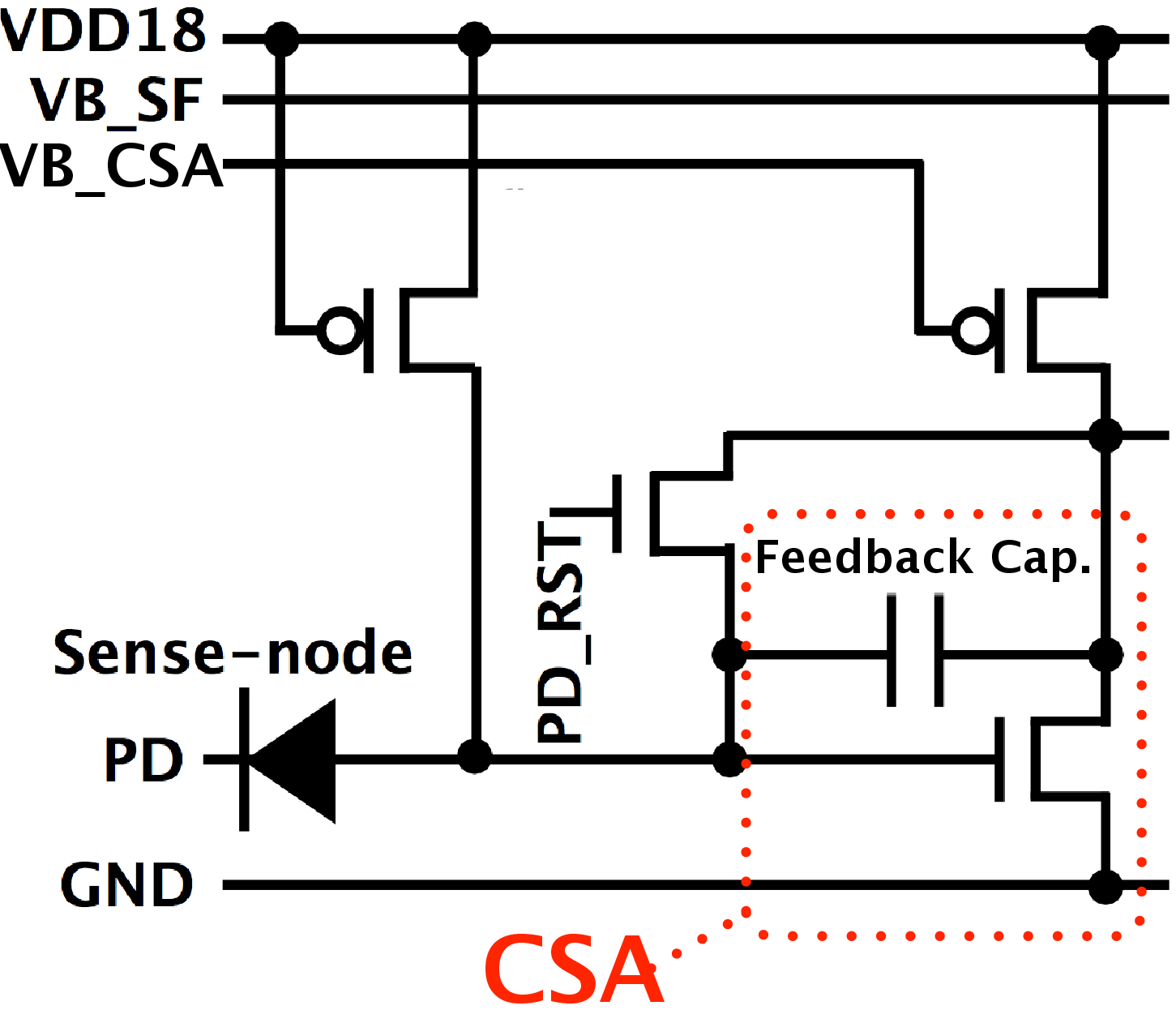}
   \end{tabular}
   \end{center}
   \caption[Figure_6]
   {\label{Figure_6}
	The circuit diagram of the in-pixel charge sensitive amplifier (CSA)
	implemented in XRPIX3\cite{2013TakedaIEEE_NSS13_XRPIX3_33e_300eV}.
	}
 \end{minipage}
\end{center}
\end{figure}

Figure~\ref{Figure_5} shows a Cd-109 spectrum obtained with XRPIX1, which is the first device developed in 2010
with a sensitivity of $3.56~{\rm \mu V~e^{-1}}$.
The energy resolution is 1.2~keV (FWHM), which is limited by the readout noise of 129~e- (rms).
This performance is far from desirable.
To increase the energy resolution, we have made two major improvements.

The first improvement is the reduction of parasitic capacitance of the sense node,
which originates in the buried p-well (BPW).
We need a thick depletion layer to detect high-energy X-rays.
Then, we apply the high back-bias voltage to increase the thickness of the depletion layer.
High back-bias voltage, on the other hand, causes problems due to the back-gate effect in the
transistor in the CMOS circuits.
Then, we create BPW in the interface between the SiO$_2$ insulator and the sensor layers
to shield the electric field from the back and suppress the back-gate
effect\cite{2011AraiNIMA_SOIPIX_STD7_BPW}.
BPW successfully improved the transistor characteristics
and allows us to apply a high back-bias voltage without the back gate effect.
Furthermore, BPW acts like part of the sense node because they are electrically connected to each other.
We found that BPW is the dominant source of parasitic capacitance of the sense node.
By reducing the BPW area by 45\%,
we successfully obtained a 1.7 times higher gain in XRPIX1b\cite{2013RyuIEEE_XRPIX1b-CZ-FI_SoftXray_CrossTalk}.

\begin{figure}
\begin{center}
 \begin{minipage}{0.45\hsize}
   \begin{center}
   \begin{tabular}{c}
   \includegraphics[height=6cm]{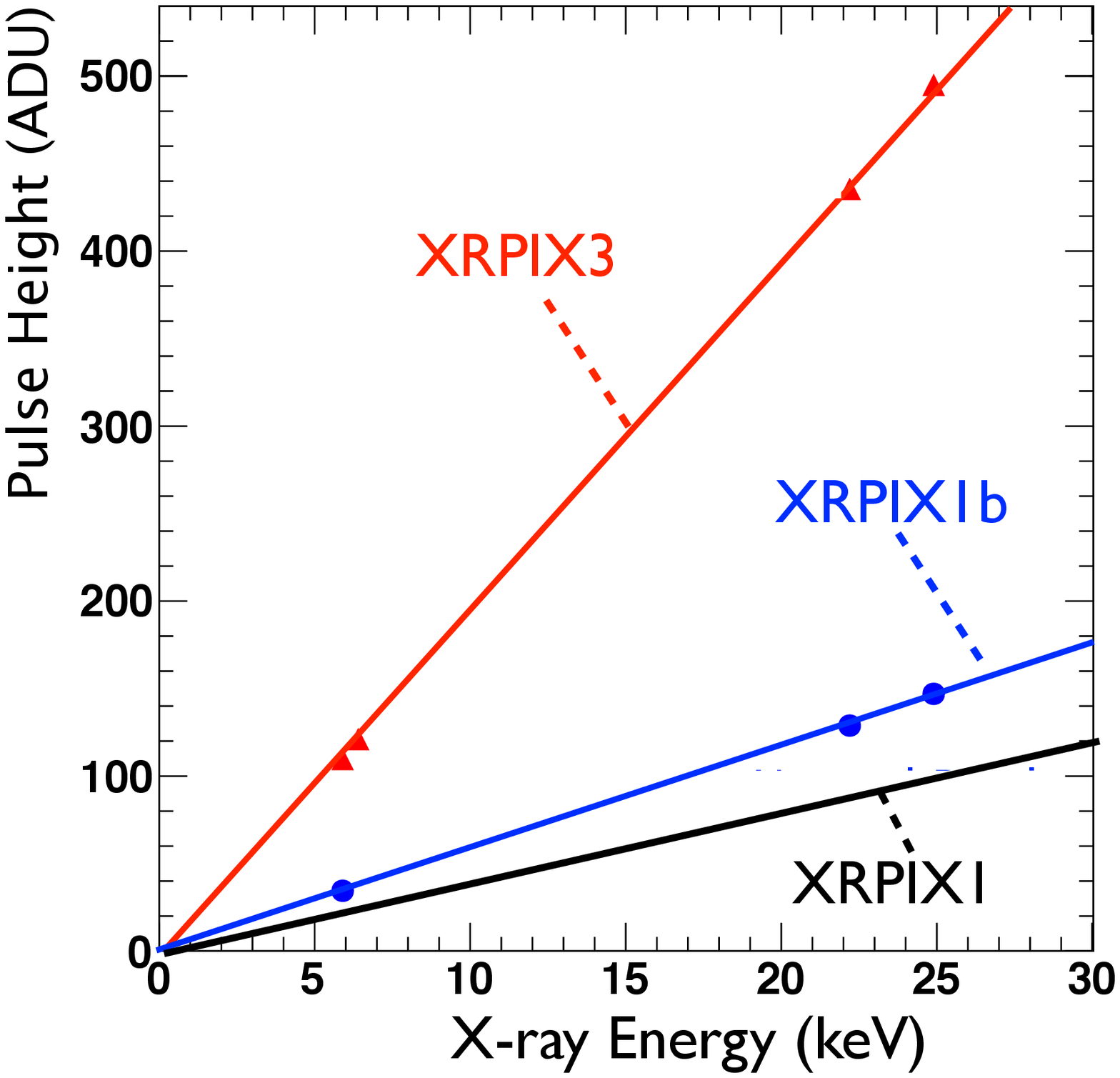}
   \end{tabular}
   \end{center}
   \caption[Figure_7]
   {\label{Figure_7}
	The peak channel as a function of the X-ray energy
	with XRPIX1 (black), XRPIX1b (blue),
	and XRPIX3 (red)\cite{2011RyuIEEE_SOIPIX_XRPIX1-CZ-FI, 2013RyuIEEE_XRPIX1b-CZ-FI_SoftXray_CrossTalk,
	2013TakedaIEEE_NSS13_XRPIX3_33e_300eV}, where
	1 ADU corresponds to $224~\mu{\rm V}$.
	}
 \end{minipage}
 \hspace{0.05\hsize}
 \begin{minipage}{0.45\hsize}
   \begin{center}
   \begin{tabular}{c}
   \includegraphics[height=6cm]{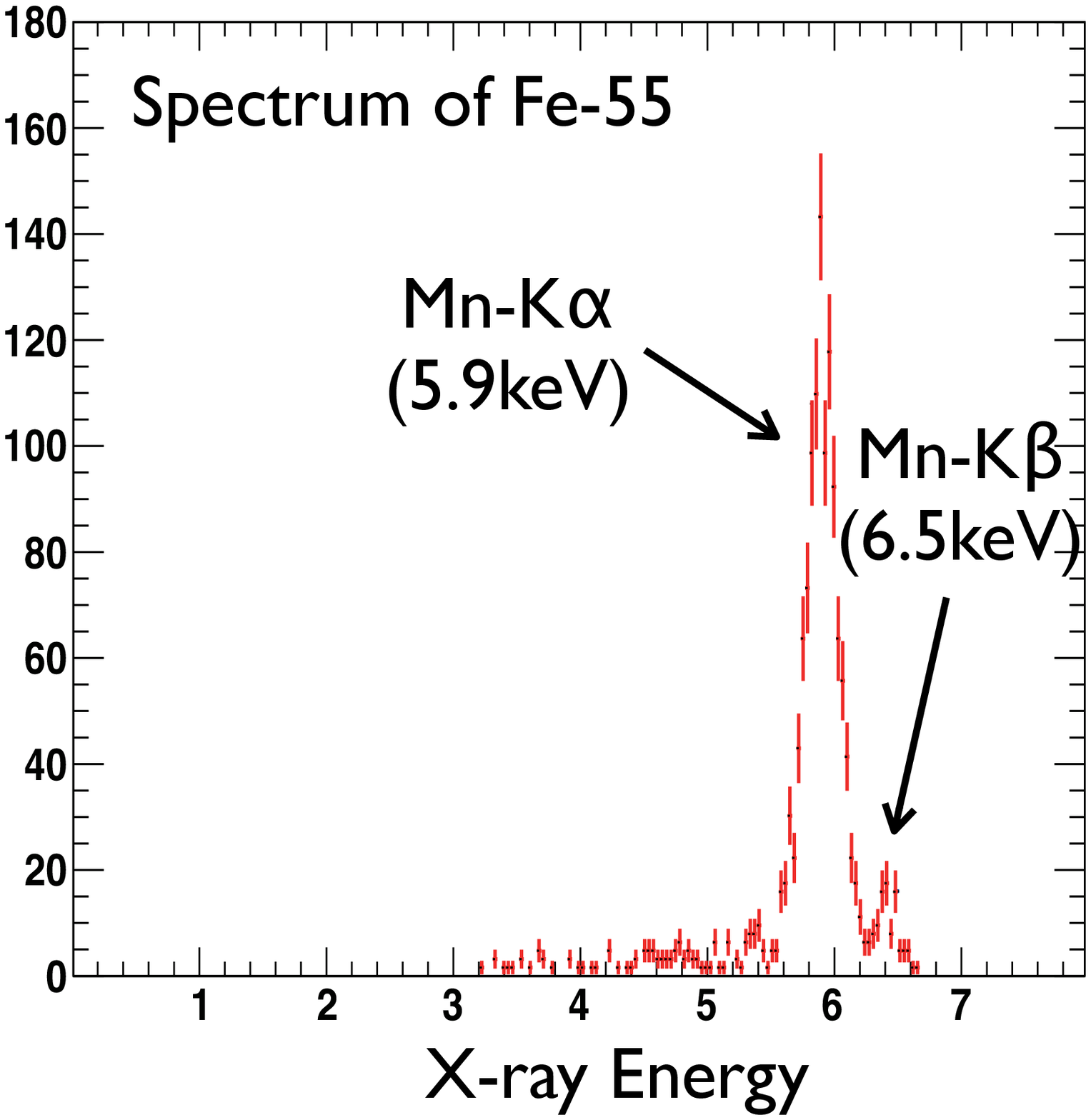}
   \end{tabular}
   \end{center}
   \caption[Figure_8]
   {\label{Figure_8}
	An Fe-55 spectrum obtained with XRPIX3-CZ\cite{2013TakedaIEEE_NSS13_XRPIX3_33e_300eV}.
	}
 \end{minipage}
\end{center}
\end{figure}

The second improvement is introduction of in-pixel charge sensitive amplifier (CSA).
Instead of the source follower for the photo-diode,
we apply a simple CSA in XRPIX3, as shown
in Figure~\ref{Figure_6} \cite{2013TakedaIEEE_NSS13_XRPIX3_33e_300eV}.
The gain was increased in XRPIX3 by factor of 5 over that of XRPIX1 (Figure~\ref{Figure_7}).

Figure~\ref{Figure_8} is an Fe-55 spectrum we obtained after making the two improvements.
We successfully resolved the K$\alpha$ and K$\beta$ lines of Mn-K X-rays and decreased the readout noise to
33~e- (rms).
Unfortunately, this design does not meet the readout noise requirements of 10~e- (rms).
Thus, we will do further improvement to the pixel and peripheral circuits.

\section{Performance in the Event-Driven Readout Mode}
\label{sec:event-driven_readout}
\begin{figure}[t]
\begin{center}
 \begin{minipage}{0.45\hsize}
   \begin{center}
   \begin{tabular}{c}
   \includegraphics[width=7cm]{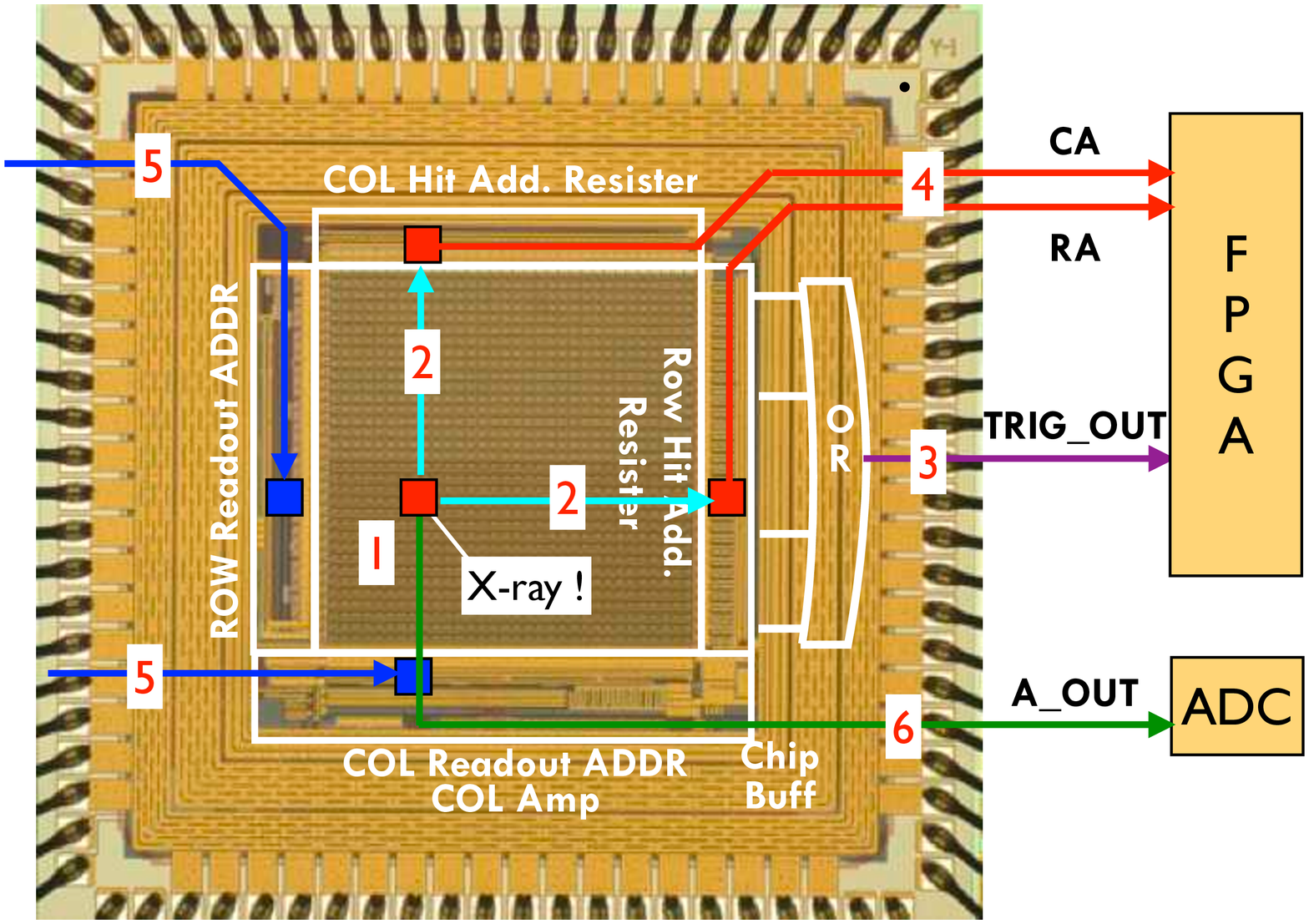}
   \end{tabular}
   \end{center}
   \caption[Figure_9]
   {\label{Figure_9}
	The readout procedure for the event-driven readout mode\cite{2011RyuIEEE-NSS11_XRPIX1_EventDriven}
(See text for details).
	The background picture is XRPIX1.
	}
 \end{minipage}
 \hspace{0.05\hsize}
 \begin{minipage}{0.45\hsize}
   \begin{center}
   \begin{tabular}{c}
   \includegraphics[width=6cm]{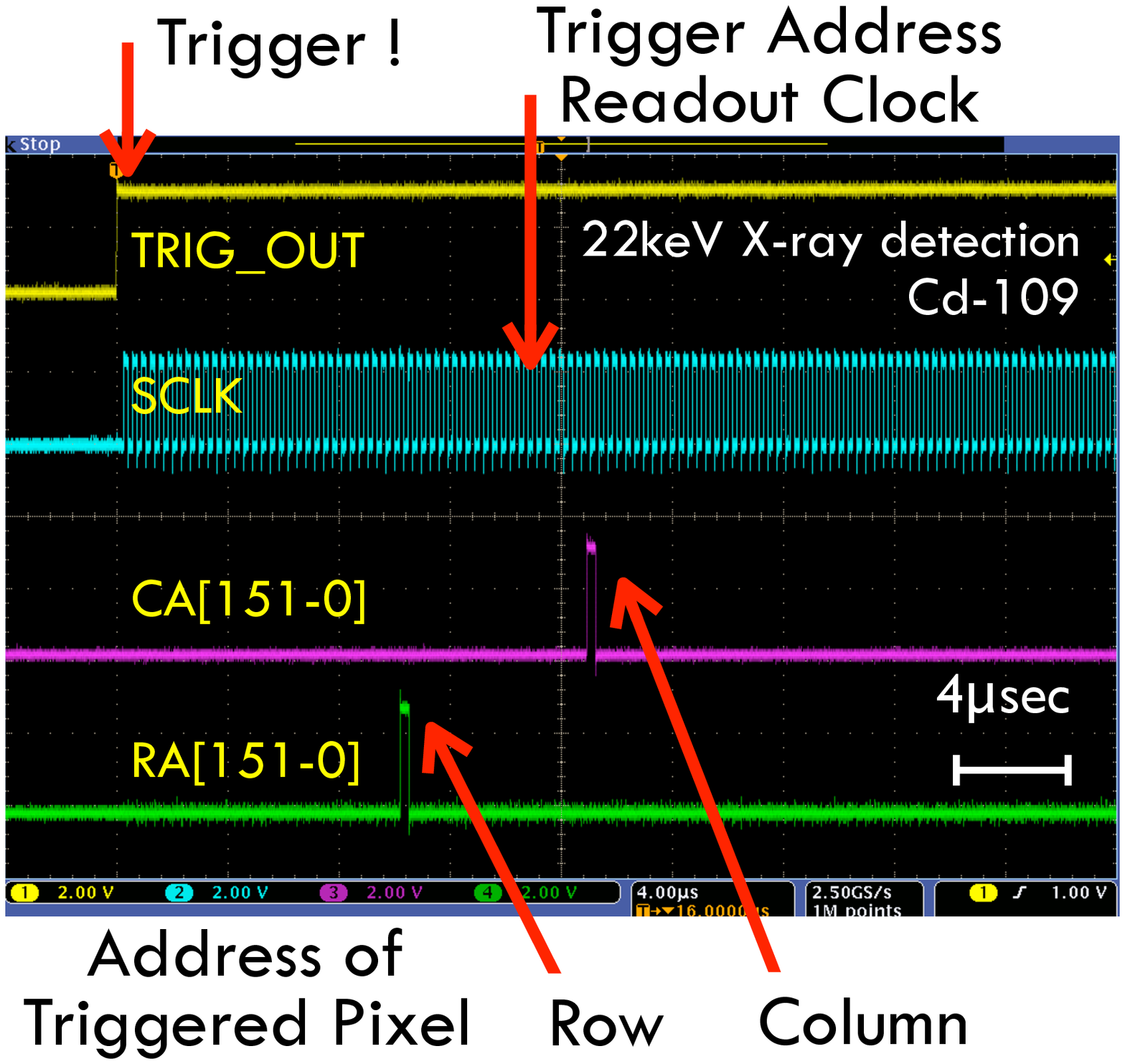}
   \end{tabular}
   \end{center}
   \caption[Figure_10]
   {\label{Figure_10}
	The waveforms of trigger out (TRIG\_OUT), the serial readout clock (SCLK), column address (CA),
	and row address (RA) for
	22-keV X-ray detection in XRPIX2b-CZ\cite{2014TakedaNIMA_XRPIX2b_EventDriven_TIPP2014}.
	}
 \end{minipage}
\end{center}
\end{figure}

Figure~\ref{Figure_9} shows the procedure from the event detection to the analog readout in the event-driven
readout mode.
(1) We assume an X-ray is absorbed at the pixel labeled ``X-ray.''
(2) The trigger circuit in the hit pixel sets its column and row addresses
in the designated serial resisters in the peripheral circuits labeled ``COL Hit Add. Resister'' and
``Row Hit Add. Resister,'' respectively.
(3) A trigger signal is immediately sent to field-programmable gate array
(FPGA) through ``TRIG\_OUT''.
(4) The FPGA reads the column and row addresses from ``Col Hit Add. Resister'' and ``Row Hit Add. Resister''
to ``CA'' and ``RA,'' respectively.
(5) The FPGA sets the column and row readout addresses
in the serial resisters ``Col Readout ADDR'' and ``Row Readout ADDR,'' respectively.
(6) Then, we finally read out the analog signal through ``COL Amp,'' ``Chip buff,'' and ``A\_OUT.''

We already successfully obtained X-ray data in the event-driven readout mode
in XRPIX1\cite{2010RyuIEEE-NSS10_XRPIX1-CZ-FI}.
Here, we present X-ray results obtained with XRPIX2b.
Figure~\ref{Figure_10} shows the waveforms of the trigger signal and other lines
when a 22-keV X-ray is detected\cite{2014TakedaNIMA_XRPIX2b_EventDriven_TIPP2014}.
The trigger speed is faster than $10~\mu {\rm sec}$,
and the capacity of the event throughput is higher than 1~kHz.

\begin{figure}
\begin{center}
 \begin{minipage}{0.45\hsize}
   \begin{center}
   \begin{tabular}{c}
   \includegraphics[height=6cm]{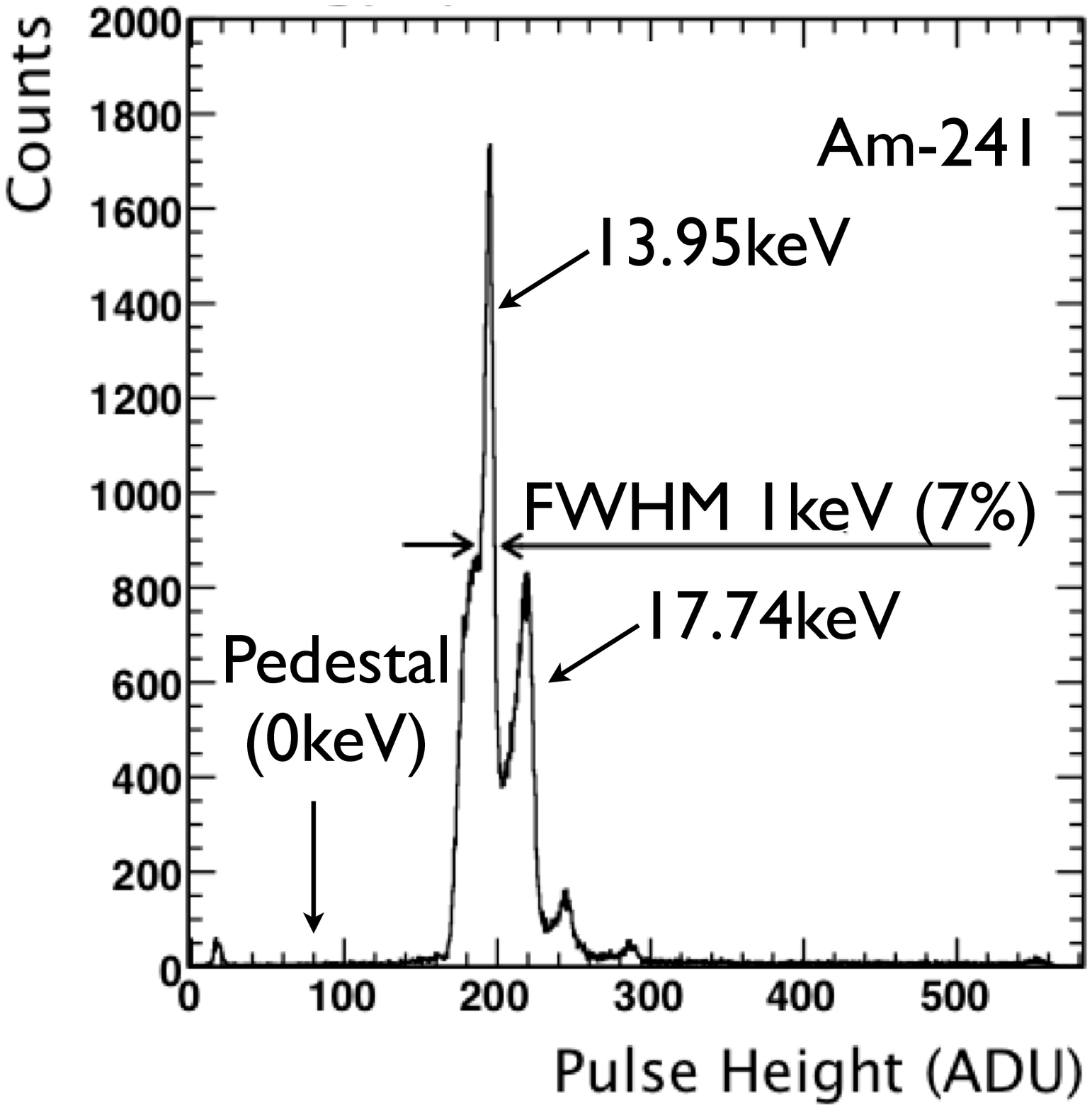}
   \end{tabular}
   \end{center}
   \caption[Figure_11]
   {\label{Figure_11}
	The Am-241 spectrum obtained with XRPIX2b-CZ in the event-driven readout mode, where
	1 ADU corresponds to $224~\mu{\rm V}$. \\ \\ \\
	}
 \end{minipage}
 \hspace{0.05\hsize}
 \begin{minipage}{0.45\hsize}
   \begin{center}
   \begin{tabular}{c}
   \includegraphics[height=6cm]{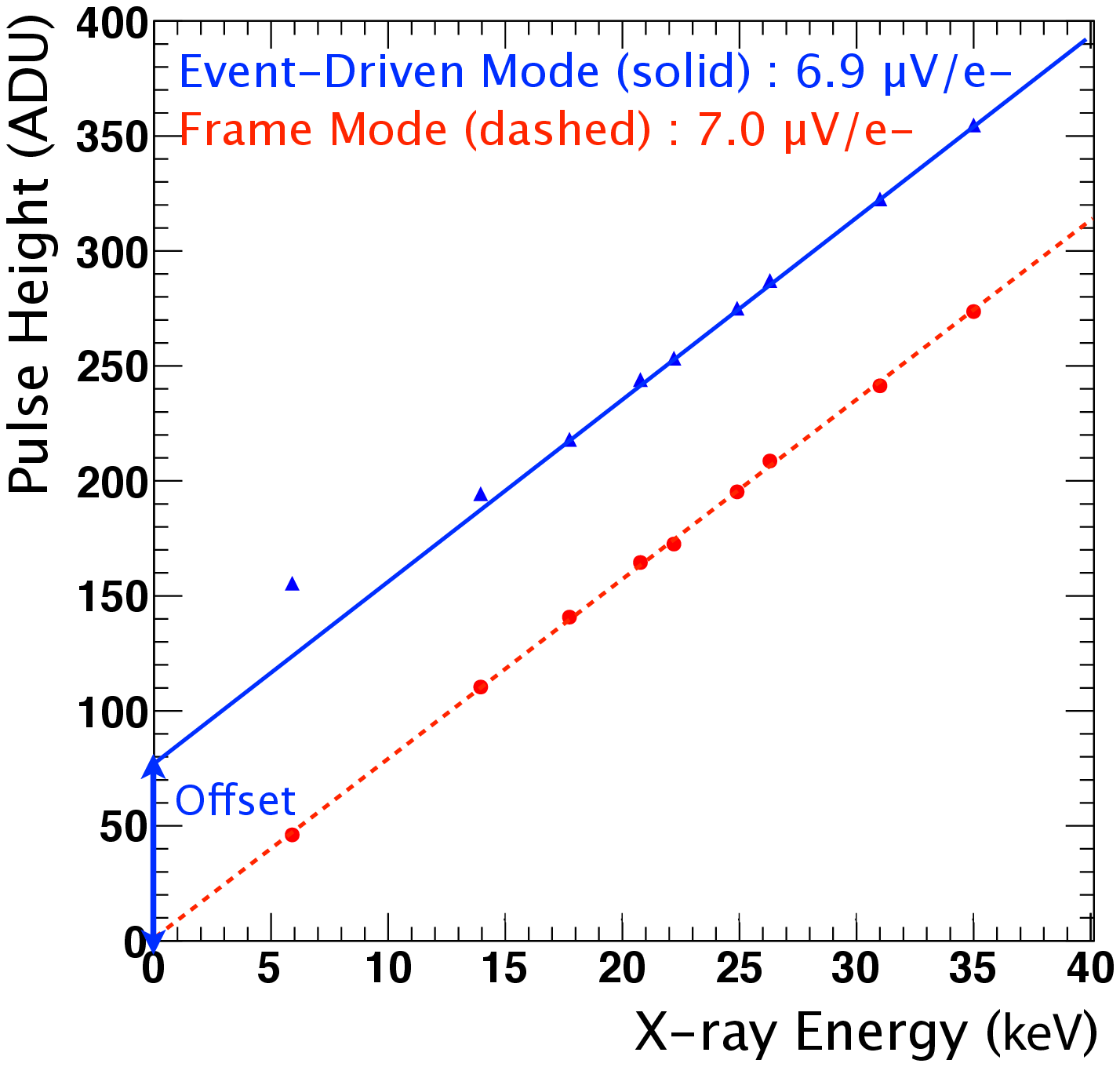}
   \end{tabular}
   \end{center}
   \caption[Figure_12]
   {\label{Figure_12}
	Peak channel as a function of the X-ray energy of XRPIX2b-CZ
	in the event-driven readout and frame readout modes, where
	1 ADU corresponds to $224~\mu{\rm V}$.
	The blue and dotted red lines are linear fits to the data of each mode.
	Note that the Fe-55 data in the event-driven readout mode is not used
	in the fitting\cite{2014TakedaNIMA_XRPIX2b_EventDriven_TIPP2014}.
	}
 \end{minipage}
\end{center}
\end{figure}

Figure~\ref{Figure_11} shows the Am-241 spectrum obtained in the event-driven readout mode. 
Energy resolution is 1000~eV in FWHM at 14~keV.
Note that XRPIX2b is a non-CSA-type pixel sensor. 
Though we successfully demonstrated the event-driven readout mode, we found some problems.
Figure~\ref{Figure_12} shows the peak channel as a function of the X-ray energy in the event-driven and
frame readout modes.
There is an offset in the pulse height of the event-driven readout mode,
even though its gain is similar to that in the frame readout mode.
Through additional experiments and simulations,
we recently determined that the offset is a result of the crosstalk between the sense node and its corresponding
circuitry and the characteristics of the inverter chopper-type comparators
used for the trigger\cite{2014TakedaNIMA_XRPIX2b_EventDriven_TIPP2014}.
We will modify the circuit layout in the upcoming device.

\section{Summary}
(1) We have been developing X-ray SOIPIX (XRPIX) for future X-ray astronomy.
(2) We reached the readout noise of 33~e- (rms) and successfully resolved Mn-K$\alpha$ and K$\beta$
with the energy resolution of $300~{\rm eV}$ (FWHM) at 6~keV in the frame readout mode.
(3) We obtained a successful readout in the event-driven readout mode and reached the energy resolution of 620~eV (FWHM)
at 6~keV with a non-CSA device, XRPIX2b-CZ.
(4) On the other hand, we found an interference problem between analog and digital circuits in the mode.
(5) We are working to further reduce the readout noise
and to mitigate the cross-talk between the sensor and circuit by modifying the layout.


\end{document}